\documentclass[amsmath,osajnl2,twocolumn,showpacs]{revtex4}
\usepackage[draft]{hyperref} %% optional
\usepackage{graphics}

\begin{document} 

\title{Nanofibers with Bragg gratings from equidistant holes}

\author{Fam Le Kien}
\email{fam@pc.uec.ac.jp}
\altaffiliation{Also at Institute of Physics, Vietnamese Academy of Science and Technology, Hanoi, Vietnam.}

\author{K. P. Nayak}

\author{K. Hakuta} 

\affiliation{Center for Photonic Innovations and
Department of Engineering Science, 
University of Electro-Communications, Chofu, Tokyo 182-8585, Japan}

\date{\today}

\begin{abstract}
We study nanofibers with Bragg gratings from equidistant holes. We calculate analytically and numerically the reflection and transmission coefficients for a single grating and also for a cavity formed by two gratings. We show that the reflection and transmission coefficients of the gratings substantially depend on the number of holes, the hole length, the hole depth, the grating period, and the light wavelength. We find that the reflection and transmission coefficients of the gratings depend on the orientation of the polarization vector of light with respect to the holes. Such a dependence is a result of the fact that the cross section of the gratings is not cylindrically symmetric. 
\end{abstract}

% REPLACE WITH CORRECT OCIS CODES FOR YOUR ARTICLE
                          % NOTE: \ocis{} IS ALIASED TO \pacs{} BUT MUST
                          % FORMAT THE TERMS CORRECTLY FOR EACH JOURNAL
\maketitle

\section{Introduction}

A fiber Bragg grating (FBG) is a structure with a periodic variation in the refractive index in a small section of an optical fiber. A FBG can be formed from multiple layers of alternating materials with varying refractive index, or by a periodic variation of some characteristic of the fiber core. The change in the refractive index in the axial direction of the fiber causes a partial reflection of an optical wave. For waves whose wavelength is close to twice the grating period, the many reflections combine with constructive interference, and the structure acts as a high-quality reflector. 
The first FBG was demonstrated by Hill \textit{et al.} using a visible laser propagating along the fiber core \cite{Hill}. Meltz \textit{et al.} demonstrated the much more flexible transverse holographic technique where the laser illumination came from the side of the fiber \cite{Meltz}. This technique uses the interference pattern of ultraviolet laser light to create the periodic structure of the Bragg grating. FBGs have been widely used in many applications \cite{Kashyap,Canning,Wan,Chow,Gupta}. The primary application of FBGs is in optical communications systems. They are specifically used as notch filters. They are also used in optical multiplexers, demultiplexers, and optical fiber sensors. 

Due to recent developments in the taper fiber technology, thin fibers can be produced with diameters smaller than the wavelength of light \cite{Mazur's Nature,Birks}. The essence of the technology is to heat and pull a single-mode optical fiber to a very small thickness, maintaining the taper condition to keep adiabatically the single-mode condition \cite{taper}. Due to tapering, the original core is almost vanishing. Therefore, the refractive indices that determine the guiding properties of the tapered fiber are the refractive index of the original silica clad and the refractive index of the surrounding vacuum. Such subwavelength-diameter vacuum-clad silica-core fibers are called nanofibers. 
Several methods for trapping and guiding neutral atoms outside a nanofiber have been proposed and studied \cite{Dowling,onecolor,twocolors,Rauschenbeutel,twocolor experiment}. A tapered fiber with an intense evanescent field can be used as an atomic mirror \cite{Bures and Ghosh}. The generation of light with a supercontinuum spectrum in thin tapered fibers has been demonstrated \cite{Birks}. The evanescent waves from zero-mode metal-clad subwavelength-diameter waveguides have been used for optical observations of single-molecule dynamics \cite{zero mode}. Efficient channeling of
emission from a few atoms into guided modes has been realized \cite{Kali}. The correlations between photons emitted from a few atoms have been measured \cite{Kali antibunching}. Thin fiber structures can be used as building blocks in future atom and photonic micro- and nanodevices. Thin fibers can also be used as efficient nanoprobes for atoms, molecules, and quantum dots.

Recently, it has been proposed to combine the FBG technique with the nanofiber technique to obtain a hybrid system \cite{fibercavity}. In such a system, two FBG mirrors form a cavity transmitting and reflecting the guided field of the nanofiber. Therefore, the interaction between the field and the atoms in the vicinity of the fiber is enhanced not only by the transverse confinement of the field in the fiber cross-section plane but also by the longitudinal confinement of the field in the space between the mirrors. An advantage of the nanofiber cavity is that the field in the guided modes can be confined to a small cross-section area whose size is comparable to the light wavelength. Another advantage of the nanofiber cavity is that the cavity guided field can be transmitted over long distances for communication purposes. The nanofiber cavity can find applications in the merged areas of fiber optics, cavity quantum electrodynamics, ultracold neutral atoms, and electromagnetically induced transparency \cite{Mabuchi,Hood Science,Aoki,Lukin1998,Xiao,Zhu}. It has been shown that a nanofiber cavity with a large length (on the order of 10 cm) and a moderate finesse (about 30) can significantly enhance the group delay of a guided probe field \cite{fibercavity} and substantially enhance the channeling of emission from an atom into the nanostructure \cite{cavityspon}. The effect of an atom on a quantum guided field in a weakly driven FBG cavity has been studied \cite{cavitytrap}. 

In order to produce a fiber cavity, one may splice commercial FBGs to a fiber. However, this technique is not suitable for making a nanofiber cavity because it creates substantial losses in the tapering region. Meanwhile, the UV irradiation technique for making FBGs uses the photosensitivity of the Ge/GeO$_2$-doped core of optical fibers. However, this technique does not work for nanofibers because the original core is almost vanishing while the original clad acts as the core of the nanofiber. Recently, a new technique for making nanofiber cavities has been demonstrated \cite{nanogrooves}. This technique uses focused ion beams \cite{FIB} to drill periodic holes on a nanofiber. The typical finesse of a nanofiber cavity produced by the focused-ion-beam milling technique is $F\sim$ 20--120 and the typical on-resonance transmission is $\sim$ 30--80\%. Such nanofiber cavities can find applications in the merged areas of fiber optics and cavity quantum electrodynamics \cite{fibercavity,Mabuchi,Hood Science,Aoki,Lukin1998,Xiao,Zhu}.

In this article, we study nanofibers with Bragg gratings from equidistant holes. We calculate analytically and numerically the reflection and transmission coefficients for a single grating and also for a cavity formed by two gratings. We find that the reflection and transmission coefficients of the gratings depend on the orientation of the polarization vector of light with respect to the holes. 

The article is organized as follows. In Sec.~\ref{sec:model} we describe the model and present the basic equations of the coupled-mode theory for the system. In Sec.~\ref{sec:grating} we calculate the reflection and transmission coefficients of a single grating. In Sec.~\ref{sec:cavity} we calculate the reflection and transmission coefficients of a cavity formed by two gratings. Our conclusions are given in Sec.~\ref{sec:summary}.

\section{Model and basic equations}
\label{sec:model}

\begin{figure}[tbh]
\begin{center}
 \includegraphics{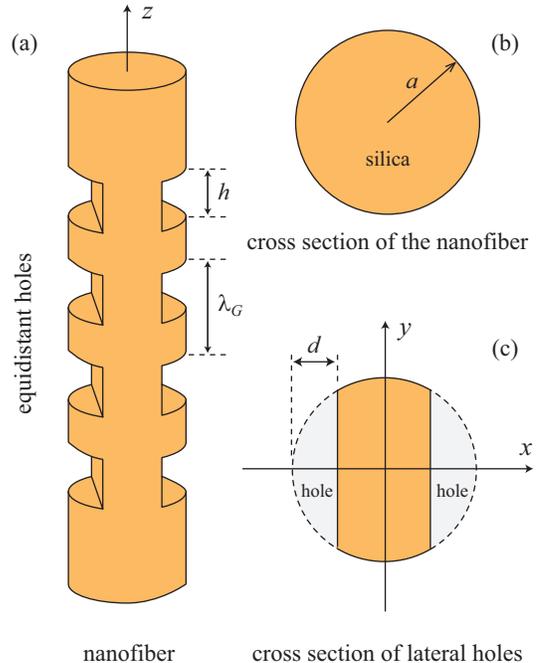}
 \end{center}
\caption{(a) Nanofiber with a grating of equidistant pairs of lateral identical holes.
(b) Cross section of the nanofiber in the transverse plane.
(c) Cross section of a pair of holes carved into the fiber on the opposite lateral sides.
}
\label{fig1}
\end{figure}

We consider a nanofiber with a grating of equidistant pairs of lateral identical holes (see Fig. \ref{fig1}). The nanofiber has a cylindrical silica core of radius $a$ and of refractive index $n_1=1.45$ and an infinite vacuum clad of refractive index $n_2=1$. In view of the very low losses of silica in the wavelength range of interest, we neglect material absorption. We use the Cartesian coordinates $(x,y,z)$ and the cylindrical coordinates $(r,\varphi,z)$, with $z$ being the axis of the fiber. 

The holes are carved on the opposite lateral sides of the fiber [see Fig. \ref{fig1}(a)]. They are paired, one left and one right. Such a grating can be produced \cite{nanogrooves} by the focused ion beam milling method \cite{FIB}. The cross section of the nanofiber in the absence of holes is a disk [see Fig. \ref{fig1}(b)]. The medium in this area is silica, with the refractive index $n_1=1.45$. The cross section of a hole has the form of a circular segment [see Fig. \ref{fig1}(c)]. The width $d$ of the circular segment is the depth of the hole. Each hole is symmetric about the axis $x$. The holes of every pair are symmetric to each other with respect to the axis $y$. The symmetry axes $x$ and $y$ and are called the principal axes. The length of a hole along the fiber axial direction $z$ is denoted by $h$. The spatial period of the grating of holes is denoted by $\lambda_G$.
Since the holes are simply vacuum voids, the refractive index of the medium in the holes is $n_2=1$.

%%%%%%%%%%%%%%%%%%%%%%% Figure 1

To describe the field, we use the basis consisting of the modes of the nanofiber in the absence of the holes. In an unperturbed (ideal) fiber, the electromagnetic field can be decomposed into guided and radiation modes \cite{fiber books}. The guided modes have the evanescent behavior on the outside of the core. They can travel in the waveguide without loss of power, provided that losses in the dielectric material are ignored. Meanwhile, the radiation modes are oscillatory at large distances from the fiber and do not have the evanescent behavior. They cannot be normalized to a finite amount of power.
  
Regarding the guided modes, we assume that the single-mode condition \cite{fiber books} is satisfied for a finite bandwidth around a central frequency of interest. We label the fundamental guided modes HE$_{11}$ in this bandwidth by the index $\mu=(\omega,f,l)$, where $\omega$ is the frequency of the mode, $f=+,-$ denotes the forward or backward propagation direction and $l=+,-$ stands for the counterclockwise or clockwise rotation of polarization. The propagation constant $\beta=\beta(\omega)$ for a guided mode is determined by the fiber eigenvalue equation \cite{fiber books}. 
The profile functions $\boldsymbol{\mathcal{E}}_{\mu}=\mathbf{e}^{(\mu)}e^{il\varphi}$ and $\boldsymbol{\mathcal{H}}_{\mu}=\mathbf{h}^{(\mu)}e^{il\varphi}$ of the electric and magnetic components, respectively, of the field in the guided mode $\mu=(\omega,f,l)$ are given in Appendix \ref{sec:guided}. 

Regarding the radiation modes, the longitudinal propagation constant $\beta$ for a given frequency $\omega$ can vary continuously, from $-k$ to $k$, with $k=\omega/c$ being the wave number. We label the radiation modes by the index $\nu=(\omega,\beta,m,l)$, where 
$m=0,\pm1,\pm2,\dots$ is the mode order and $l=+,-$ is the mode polarization. The profile functions $\boldsymbol{\mathcal{E}}_{\nu}=\mathbf{e}^{(\nu)}e^{im\varphi}$ and $\boldsymbol{\mathcal{H}}_{\nu}=\mathbf{h}^{(\nu)}e^{im\varphi}$ of the electric and magnetic components, respectively, of the field in the radiation mode $\nu=(\omega,\beta,m,l)$ are given in \cite{cavityspon,fiber books}. 

We assume that the field has a single frequency $\omega$. 
The orthogonality of the modes of the unperturbed fiber is expressed by the formula 
\begin{equation}
\frac{1}{2} \int \mathrm{Re}[\mathbf{u}_z\cdot (\boldsymbol{\mathcal{E}}_{i}\times \boldsymbol{\mathcal{H}}_{i'}^*)] \,d\mathbf{r}=P_{i}\delta_{ii'}.
\label{f1}
\end{equation}
Here $\mathbf{u}_z$ is the unit vector for the direction $z$, and $\delta_{ii'}$ is the Kronecker delta symbol for discrete indices $i$ and $i'$ of guided modes, the Dirac delta function for continuous indices $i$ and $i'$ of radiation modes, and zero if one of the indices $i$ and $i'$ is discrete and the other one is continuous. We have used the notation $\int d\mathbf{r}=\int_0^{2\pi}d\varphi\int_0^{\infty}r\,dr$ for the integral over the fiber transverse plane. 
The notation $P_{i}$ stands for the power of the mode $i$. For convenience, we normalize the mode functions to the same power $P$, that is, we choose $P_{i}=P$. 

We now consider the field in the presence of the holes. It is our purpose to expand the field in terms of the modes of the ideal fiber. We separate the field vectors into the transverse and longitudinal parts designated by the subscripts $t$ and $z$ as
\begin{eqnarray}
\mathbf{E}&=&\mathbf{E}_t+\mathbf{E}_z,\nonumber\\
\mathbf{H}&=&\mathbf{H}_t+\mathbf{H}_z.
\label{f2}
\end{eqnarray}
Due to the completeness of the set of transverse mode functions, we can expand the transverse components $\mathbf{E}_t$ and $\mathbf{H}_t$ of the electric and magnetic parts of the field in terms of the ideal modes as
\begin{eqnarray}
\mathbf{E}_t=\sum_{\alpha}(a_{\alpha}^++a_{\alpha}^-)\boldsymbol{\mathcal{E}}_{\alpha t},\nonumber\\
\mathbf{H}_t=\sum_{\alpha}(a_{\alpha}^+-a_{\alpha}^-)\boldsymbol{\mathcal{H}}_{\alpha t}.
\label{f3}
\end{eqnarray}
Here $a_{\alpha}^+$ and $a_{\alpha}^-$ are the amplitudes of the field in the forward and backward modes, respectively, $\boldsymbol{\mathcal{E}}_{\alpha}$ and $\boldsymbol{\mathcal{H}}_{\alpha}$
are the profile functions for the electric and magnetic parts of the forward modes, 
and $\alpha=\mu,\nu$ is the common label for the guided and radiation modes. The longitudinal components $E_z$ and $H_z$ of the electric and magnetic parts of the field can be written as 
\begin{eqnarray}
\mathbf{E}_z=\frac{n_0^2}{n^2}\sum_{\alpha}(a_{\alpha}^+-a_{\alpha}^-)\boldsymbol{\mathcal{E}}_{\alpha z},\nonumber\\
\mathbf{H}_z=\sum_{\alpha}(a_{\alpha}^++a_{\alpha}^-)\boldsymbol{\mathcal{H}}_{\alpha z},
\label{f4}
\end{eqnarray}
where the functions $n=n(x,y,z)=n(r,\varphi,z)$ and $n_0=n_0(x,y)=n_0(r,\varphi)$ are the refractive-index distributions of the perturbed and unperturbed fibers, respectively.

According to the coupled-mode theory \cite{fiber books},
the amplitudes $a_{\alpha}^+$ and $a_{\alpha}^-$ are governed by the equations
\begin{eqnarray}
\frac{d a_{\alpha}^+}{d z}&=&i\beta_{\alpha}a_{\alpha}^+ +i\sum_{\alpha'} (U_{\alpha\alpha'}a_{\alpha'}^+ +
V_{\alpha\alpha'}a_{\alpha'}^-),\nonumber\\
\frac{d a_{\alpha}^-}{d z}&=&-i\beta_{\alpha}a_{\alpha}^- -i\sum_{\alpha'}(V_{\alpha\alpha'}a_{\alpha'}^+ +
U_{\alpha\alpha'}a_{\alpha'}^-),
\label{f5}
\end{eqnarray}
where 
\begin{eqnarray}
U_{\alpha\alpha'}&=&\frac{\omega\epsilon_0}{4P}\int
(n^2-n_0^2)[(\boldsymbol{\mathcal{E}}_{\alpha t}\cdot \boldsymbol{\mathcal{E}}_{\alpha' t}^* )\nonumber\\&&\mbox{}
+(n_0^2/n^2)(\boldsymbol{\mathcal{E}}_{\alpha z}\cdot \boldsymbol{\mathcal{E}}_{\alpha' z}^* )
] d\mathbf{r},
\nonumber\\
V_{\alpha\alpha'}&=&\frac{\omega\epsilon_0}{4P}\int
(n^2-n_0^2)[(\boldsymbol{\mathcal{E}}_{\alpha t}\cdot \boldsymbol{\mathcal{E}}_{\alpha' t}^* )\nonumber\\&&\mbox{}
-(n_0^2/n^2)(\boldsymbol{\mathcal{E}}_{\alpha z}\cdot \boldsymbol{\mathcal{E}}_{\alpha' z}^* )
] d\mathbf{r}
\label{f6}
\end{eqnarray}
are the mode-coupling coefficients. We note that the coupled-mode equations (\ref{f5}) involve the complete set of guided and radiation modes and are exact.

We now assume that the coupling between the guided modes and the radiation modes is negligible, that is, the grating of the holes along the fiber is lossless. When we omit the radiation modes from Eqs. (\ref{f5}), we obtain the following propagation equations for the coupled guided modes:
\begin{eqnarray}
\frac{d a_{l}^+}{d z}&=&i\beta_{l}a_{l}^+ 
+i\sum_{l'} (U_{ll'}a_{l'}^+ + V_{ll'}a_{l'}^-),
\nonumber\\
\frac{d a_{l}^-}{d z}&=&-i\beta_{l}a_{l}^- 
-i\sum_{l'}(V_{ll'}a_{l'}^+ + U_{ll'}a_{l'}^-).
\label{f7}
\end{eqnarray}
Here $l$ is the guided-mode index and takes the value $1$ or $-1$, corresponding to the left or right circular polarization, respectively. We emphasize that the omission of the radiation modes from the coupled-mode equations is the only approximation used in this paper. This approximation is valid only when the losses are not serious, that is, when the characteristic size of the holes is small as compared to the fiber radius. It is interesting to note that, according to \cite{nanogrooves}, the losses are not too serious (20--70\%) even when the length and depth of the holes are substantial fractions of the fiber radius. For such losses, our analytical and numerical results shown below are not quantitatively valid. However, they can still be used to qualitatively understand the underlying physics and to explain many features of the nanofiber cavities produced by the focused-ion-beam milling method \cite{nanogrooves}.

Note that $\beta_1=\beta_{-1}\equiv\beta$.
Due to the symmetry of the mode functions and the hole shape, we have $U_{1,1}=U_{-1,-1}\equiv U$, $V_{1,1}=V_{-1,-1}\equiv V$, $U_{1,-1}=U_{-1,1}\equiv U_c$, and $V_{1,-1}=V_{-1,1}\equiv V_c$. In addition, we can show that $U=U^*$, $V=V^*$, $U_c=U_c^*$, and $V_c=V_c^*$. 
Hence, Eqs. (\ref{f7}) become
\begin{eqnarray}
\frac{d a_{1}^+}{d z}&=&i(\beta+U) a_{1}^+ + i Va_{1}^- + i U_ca_{-1}^+ + i V_ca_{-1}^-,
\nonumber\\
\frac{d a_{-1}^+}{d z}&=&i(\beta+U) a_{-1}^+ + i Va_{-1}^- + i U_ca_{1}^+ + i V_ca_{1}^-,
\nonumber\\
\frac{d a_{1}^-}{d z}&=&-i(\beta+U) a_{1}^- - iVa_{1}^+ -i U_ca_{-1}^- - iV_ca_{-1}^+,
\nonumber\\
\frac{d a_{-1}^-}{d z}&=&-i(\beta+U) a_{-1}^- - iVa_{-1}^+ -i U_ca_{1}^- - iV_ca_{1}^+.
\label{f8}
\end{eqnarray}
We note that the coefficients $U_c$ and $V_c$ characterize the coupling between the two different circular polarizations $l=\pm1$. Such cross coupling arises because the cylindrical symmetry of the fiber is broken by the presence of the holes on the lateral sides.

We introduce the notations $a_x=(a_{1}+a_{-1})/\sqrt2$ and $a_y=(a_{1}-a_{-1})/i\sqrt2$,
which stand for the amplitudes of the fields that are linearly polarized along the $x$ and $y$ directions.
Then, Eqs. (\ref{f8}) split into two different sets of coupled equations, namely
\begin{eqnarray}
\frac{d a_{x}^+}{d z}&=&i(\beta+U_x) a_{x}^+ + iV_xa_{x}^-,
\nonumber\\
\frac{d a_{x}^-}{d z}&=&-i(\beta+U_x) a_{x}^- - iV_xa_{x}^+,
\label{f9}
\end{eqnarray}
and
\begin{eqnarray}
\frac{d a_{y}^+}{d z}&=&i(\beta+U_y) a_{y}^+ + iV_ya_{y}^-,
\nonumber\\
\frac{d a_{y}^-}{d z}&=&-i(\beta+U_y) a_{y}^- - iV_ya_{y}^+.
\label{f10}
\end{eqnarray}
Here we have introduced the notations
\begin{eqnarray}
U_x&=&U+U_c,
\nonumber\\
V_x&=&V+V_c,
\nonumber\\
U_y&=&U-U_c,
\nonumber\\
V_y&=&V-V_c.
\label{f11}
\end{eqnarray}
Equations (\ref{f9}) and (\ref{f10}) show that the $x$ and $y$ polarizations are not coupled to each other. Consequently, the $x$ and $y$ polarizations can be considered as the principal polarizations of the field in the nanofiber in the presence of the holes.
Equations (\ref{f11}) show that the principal polarizations $x$ and $y$ have different coupling coefficients.

\section{Single gratings}
\label{sec:grating}

In this section, we calculate the reflection and transmission coefficients of a single grating. 
We start our calculations by considering a single step of a grating, that is, a single pair of holes. We assume that the two holes stretch from the point $z=0$ to the point $z=h$ along the fiber axis. In the interval $(0,h)$, the coupling coefficients $U_{\sigma}$ and $V_{\sigma}$, where $\sigma=x,y$, are constant. For the boundary condition $a_{\sigma}^-(h)=0$, the solutions to Eqs. (\ref{f9}) and (\ref{f10}) are found to be
\begin{eqnarray}
a_{\sigma}^+(z)&=&A_{\sigma}\{K_{\sigma}\cos[K_{\sigma} (z-h)]
\nonumber\\&&\mbox{}
+i(\beta+U_{\sigma})\sin [K_{\sigma} (z-h)]\},
\nonumber\\
a_{\sigma}^-(z)&=&-iA_{\sigma}V_{\sigma}\sin[K_{\sigma} (z-h)],
\label{f12}
\end{eqnarray}
where
\begin{equation}
K_{\sigma}=\sqrt{(\beta+U_{\sigma})^2-V_{\sigma}^2}.
\label{f13}
\end{equation}
The coefficient $A_{\sigma}$ is determined by the input field amplitude $a_{\sigma}^+(0)$.

From Eqs. (\ref{f12}) and (\ref{f13}), the reflection coefficient $R_{\sigma}=a_{\sigma}^-(0)/a_{\sigma}^+(0)$ and the transmission coefficient $T_{\sigma}=a_{\sigma}^+(h)/a_{\sigma}^+(0)$ of the single pair of holes for the field with the principal polarization $\sigma=x,y$ are found to be
\begin{eqnarray}
R_{\sigma}&=&\frac{iV_{\sigma}\sin (K_{\sigma} h)}{K_{\sigma}\cos (K_{\sigma} h)-i(\beta+U_{\sigma})\sin (K_{\sigma} h)},
\nonumber\\
T_{\sigma}&=&\frac{K_{\sigma}}{K_{\sigma}\cos (K_{\sigma} h)-i(\beta+U_{\sigma})\sin (K_{\sigma} h)}.
\label{f14}
\end{eqnarray}
It is clear that the reflection and transmission coefficients of the pair of holes depend on the orientation of the polarization vector of the field. The reason is that, when the total cross section of the pair of holes is not cylindrically symmetric, the cross-coupling coefficients $U_c$ and $V_c$ become nonzero. Therefore, we have $U_x\not=U_y$ and $V_x\not=V_y$ [see Eqs. (\ref{f11})], and hence $K_x\not=K_y$ [see Eq. (\ref{f13})]. This leads to $R_x\not=R_y$ and $T_x\not=T_y$ [see Eqs. (\ref{f14})].

We now consider a grating of $N$ equidistant identical pairs of holes stretching from the position $z=0$ to the position $z=L_G=Nh+(N-1)s$. Here $s=\lambda_G-h$ is the length of the region that separates two adjacent pairs of holes. We consider a principal polarization $\sigma=x,y$. 
The solution of Eqs. (\ref{f9}) or (\ref{f10}) for the first step of the grating (the first pair of holes) can be written as
\begin{equation}
\left(\begin{array}{c}a_{\sigma}^{+}(h)\\a_{\sigma}^{-}(h)\end{array}\right)
=\mathbf{M}\left(\begin{array}{c}a_{\sigma}^{+}(0)\\a_{\sigma}^{-}(0)\end{array}\right),
\label{f15}
\end{equation}
where
\begin{equation}
\mathbf{M}=
\left(\begin{array}{cc}M_{11}&M_{12}\\M_{21}&M_{22}\end{array}\right)
\label{f16}
\end{equation}
is the transfer matrix for a single step, with the matrix elements
\begin{eqnarray}
M_{11}&=&\cos(K_{\sigma}h)+i\frac{\beta+U_{\sigma}}{K_{\sigma}}\sin(K_{\sigma}h),\nonumber\\
M_{22}&=&\cos(K_{\sigma}h)-i\frac{\beta+U_{\sigma}}{K_{\sigma}}\sin(K_{\sigma}h),\nonumber\\
M_{12}&=&i\frac{V_{\sigma}}{K_{\sigma}}\sin(K_{\sigma}h),\nonumber\\
M_{21}&=&-i\frac{V_{\sigma}}{K_{\sigma}}\sin(K_{\sigma}h).
\label{f17}
\end{eqnarray}
Note that $M_{12}=-M_{21}$ and $M_{11}M_{22}-M_{12}M_{21}=1$.

In the region from $z=h$ to $z=h+s=\lambda_G$, which separates the first and second steps, the fiber is unperturbed and, hence, the field propagates as an ideal guided-mode field with the propagation constant $\beta$.
Therefore, we have
\begin{equation}
\left(\begin{array}{c}a_{\sigma}^{+}(h+s)\\a_{\sigma}^{-}(h+s)\end{array}\right)
=\mathbf{F}\left(\begin{array}{c}a_{\sigma}^{+}(h)\\a_{\sigma}^{-}(h)\end{array}\right),
\label{f18}
\end{equation}
where
\begin{equation}
\mathbf{F}=
\left(\begin{array}{cc}e^{i\beta s}&0\\0&e^{-i\beta s}\end{array}\right)
\label{f19}
\end{equation}
is the propagator for the ideal guided modes of the fiber.

When we apply the above procedures to the $N$ steps of the grating and combine the results, we find the input-output relation
\begin{equation}
\left(\begin{array}{c}a_{\sigma}^{+}(L)\\a_{\sigma}^{-}(L)\end{array}\right)
=\mathbf{W}^{(N)}\left(\begin{array}{c}a_{\sigma}^{+}(0)\\a_{\sigma}^{-}(0)\end{array}\right),
\label{f20}
\end{equation}
where
\begin{equation}
\mathbf{W}^{(N)}=\mathbf{M}(\mathbf{F}\mathbf{M})^{N-1}
\label{f21}
\end{equation}
is the total transfer matrix for the grating. We note that
$\det \mathbf{W}^{(N)}=W^{(N)}_{11}W^{(N)}_{22}-W^{(N)}_{12}W^{(N)}_{21} =1$ and $W^{(N)}_{12}=-W^{(N)}_{21}$. In addition, we have $\mathbf{W}^{(N=1)}=\mathbf{M}$. The explicit expressions for the elements of the total transfer matrix $\mathbf{W}^{(N)}$ are found to be 
\begin{eqnarray}
W^{(N)}_{11}&=&M_{11}\frac{\sinh(N\theta_{\sigma})}{\sinh\theta_{\sigma}}
-e^{-i\beta s}\frac{\sinh[(N-1)\theta_{\sigma}]}{\sinh\theta_{\sigma}},
\nonumber\\
W^{(N)}_{22}&=&M_{22}\frac{\sinh(N\theta_{\sigma})}{\sinh\theta_{\sigma}}
-e^{i\beta s}\frac{\sinh[(N-1)\theta_{\sigma}]}{\sinh\theta_{\sigma}},
\nonumber\\
W^{(N)}_{12}&=&M_{12}\frac{\sinh(N\theta_{\sigma})}{\sinh\theta_{\sigma}},
\nonumber\\
W^{(N)}_{21}&=&M_{21}\frac{\sinh(N\theta_{\sigma})}{\sinh\theta_{\sigma}},
\label{f22}
\end{eqnarray}
where 
\begin{equation}
\theta_{\sigma}=\ln (D_{\sigma}+\sqrt{D_{\sigma}^2-1}), 
\label{f23}
\end{equation}
with $D_{\sigma}=\frac{1}{2}\big(M_{11}e^{i\beta s}+M_{22}e^{-i\beta s}\big)$, that is, 
\begin{equation}
D_{\sigma}=\cos(K_{\sigma}h)\cos(\beta s)-\frac{\beta+U_{\sigma}}{K_{\sigma}}\sin(K_{\sigma}h)\sin(\beta s).
\label{f24}
\end{equation}
Note that $\theta_{\sigma}$ is, in general, a complex number.

The reflection and transmission coefficients of the grating are given by
$R_{\sigma}^{(N)}=W^{(N)}_{12}/W^{(N)}_{22}=-W^{(N)}_{21}/W^{(N)}_{22}$ and $T_{\sigma}^{(N)}=1/W^{(N)}_{22}$, respectively. The explicit expressions for these coefficients are found to be
\begin{eqnarray}
R_{\sigma}^{(N)}&=&\frac{R_{\sigma}\sinh(N\theta_{\sigma})}{\sinh(N\theta_{\sigma})-T_{\sigma}e^{i\beta s}\sinh[(N-1)\theta_{\sigma}]},
\nonumber\\
T_{\sigma}^{(N)}&=&\frac{T_{\sigma}\sinh\theta_{\sigma}}{\sinh(N\theta_{\sigma})-T_{\sigma}e^{i\beta s}\sinh[(N-1)\theta_{\sigma}]}.
\label{f25}
\end{eqnarray}
Here $R_{\sigma}=R_{\sigma}^{(N=1)}=M_{12}/M_{22}=-M_{21}/M_{22}$ and $T_{\sigma}=T_{\sigma}^{(N=1)}=1/M_{22}$ are the reflection and transmission coefficients for a single step and are given by Eqs. (\ref{f14}). Equations (\ref{f25}) show that, like the coefficients $R_{\sigma}$ and $T_{\sigma}$, the coefficients $R_{\sigma}^{(N)}$ and $T_{\sigma}^{(N)}$ depend on the principal polarization $\sigma=x,y$ of the field. In addition, since $\theta_{\sigma}$ is, in general, a complex number, $R_{\sigma}^{(N)}$ and $T_{\sigma}^{(N)}$ may oscillate when a parameter, such as the number of holes $N$, the grating period $\lambda_G$, the hole depth $d$, the hole length $h$, and the wavelength of light $\lambda$, varies.

We note that, when we use Eq. (\ref{f21}), we can derive the recurrence formulae
\begin{eqnarray}
R_{\sigma}^{(N+1)}&=&R_{\sigma}^{(N)}+\frac{T_{\sigma}^{(N)2}R_{\sigma}e^{2i\beta s}}{1-R_{\sigma}^{(N)}R_{\sigma}e^{2i\beta s}},
\nonumber\\
T_{\sigma}^{(N+1)}&=&\frac{T_{\sigma}^{(N)} T_{\sigma}e^{i\beta s}}{1-R_{\sigma}^{(N)}R_{\sigma}e^{2i\beta s}}.
\label{f26}
\end{eqnarray}
We can also obtain these formulae in the framework of ray optics. 
The explicit expressions (\ref{f25}) for $R_{\sigma}^{(N)}$ and $T_{\sigma}^{(N)}$ are in agreement with the recurrence formulae (\ref{f26}).

\begin{figure}[tbh]
\begin{center}
 \includegraphics{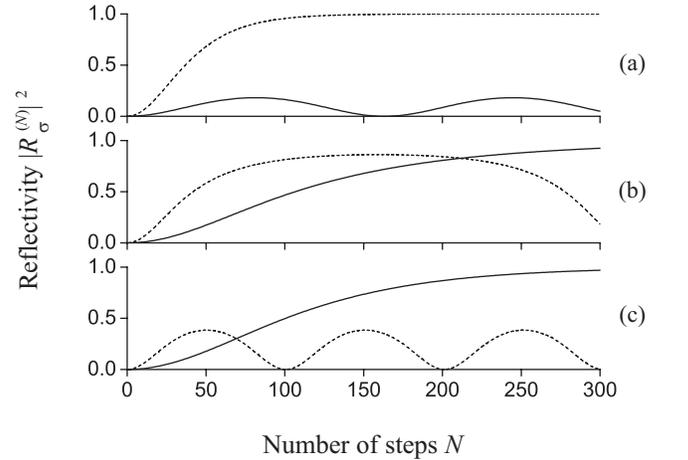}
 \end{center}
\caption{Reflectivity $|R_{\sigma}^{(N)}|^2$ as a function of the number of steps $N$. The solid and dashed curves are for the principal $x$ and $y$ polarizations, respectively. The wavelength of light is $\lambda=852$ nm. The grating period is $\lambda_G=363$ nm (a), 364.5 nm (b), and 366 nm (c). The fiber radius is $a=290$ nm. The hole length and hole depth are $h=150$ nm and $d=100$ nm, respectively. 
}
\label{fig2}
\end{figure}

\begin{figure}[tbh]
\begin{center}
 \includegraphics{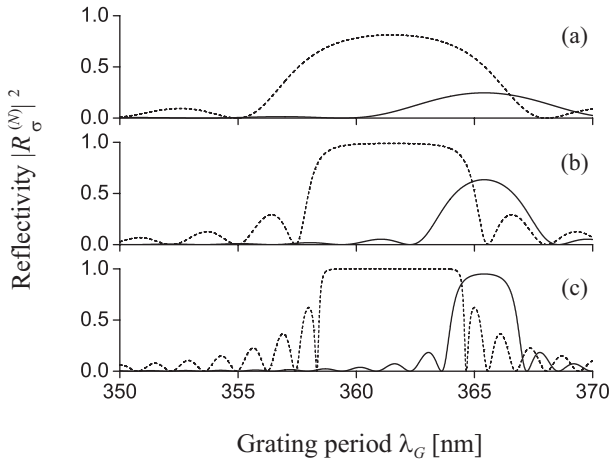}
 \end{center}
\caption{Reflectivity $|R_{\sigma}^{(N)}|^2$ as a function of the grating period $\lambda_G$. The solid and dashed curves are for the principal $x$ and $y$ polarizations, respectively. The number of steps is $N=60$ (a), 120 (b), and 240 (c). Other parameters are as in Fig. \ref{fig2}.
}
\label{fig3}
\end{figure}

\begin{figure}[tbh]
\begin{center}
 \includegraphics{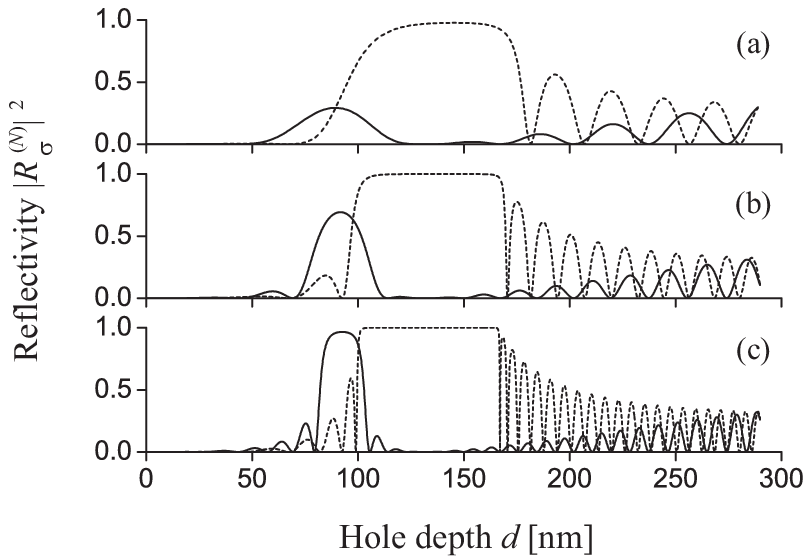}
 \end{center}
\caption{Reflectivity $|R_{\sigma}^{(N)}|^2$ as a function of the hole depth $d$. The solid and dashed curves are for the principal $x$ and $y$ polarizations, respectively. The number of steps is $N=60$ (a), 120 (b), and 240 (c). The grating period is $\lambda_G$ =364.5 nm. Other parameters are as in Fig. \ref{fig2}.
}
\label{fig4}
\end{figure}

\begin{figure}[tbh]
\begin{center}
 \includegraphics{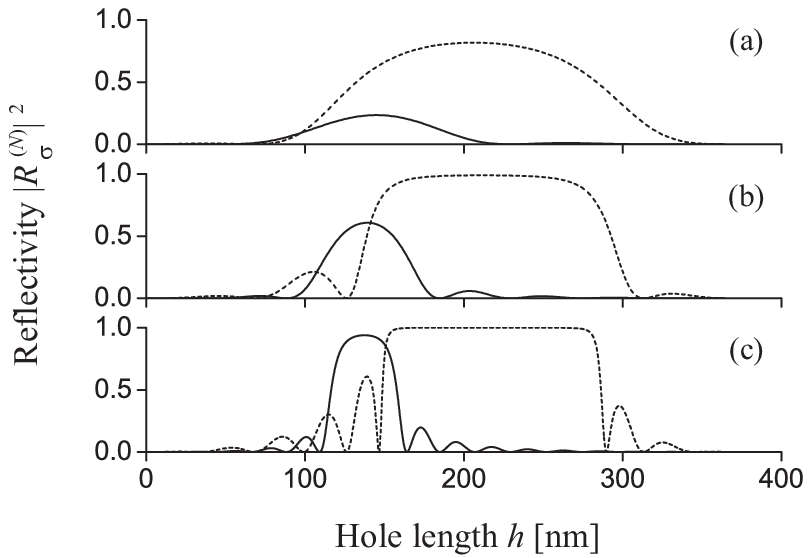}
 \end{center}
\caption{Reflectivity $|R_{\sigma}^{(N)}|^2$ as a function of the hole length $h$. The solid and dashed curves are for the principal $x$ and $y$ polarizations, respectively. The number of steps is $N=60$ (a), 120 (b), and 240 (c). The grating period is $\lambda_G$ =364.5 nm. Other parameters are as in Fig.~\ref{fig2}.
}
\label{fig5}
\end{figure}

\begin{figure}[tbh]
\begin{center}
 \includegraphics{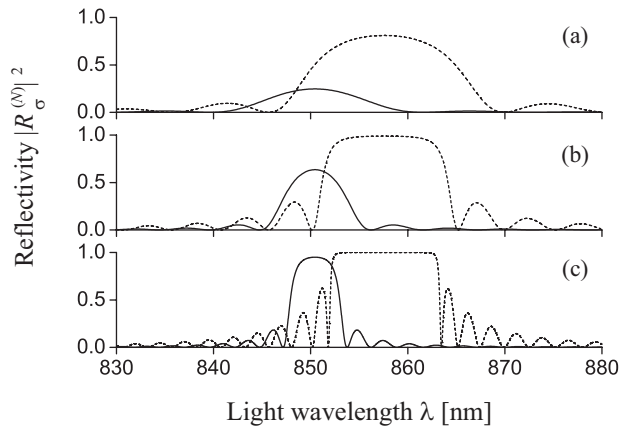}
 \end{center}
\caption{Reflectivity $|R_{\sigma}^{(N)}|^2$ as a function of the wavelength of light $\lambda$. The solid and dashed curves are for the principal $x$ and $y$ polarizations, respectively. The number of steps is $N=60$ (a), 120 (b), and 240 (c). The grating period is $\lambda_G$ =364.5 nm. Other parameters are as in Fig. \ref{fig2}.
}
\label{fig6}
\end{figure}

We calculate the reflection and transmission coefficients $R_{\sigma}^{(N)}$ and $T_{\sigma}^{(N)}$ from expressions (\ref{f25}). We plot in Figs. \ref{fig2}--\ref{fig6} the reflectivity $|R_{\sigma}^{(N)}|^2$ as a function of the number of steps $N$, the grating period $\lambda_G$, the hole depth $d$, the hole length $h$, and the wavelength of light $\lambda$. Comparison between the solid curves (for the $x$ polarization) and the dashed curves (for the $y$ polarization) shows that the different principal $x$ and $y$ polarizations of light lead to different reflectivities of the grating. 

%%%%%%%%%%%%%%%%%%%%%%% Figure 2

We observe from Fig. \ref{fig2} that the dependence of $|R_{\sigma}^{(N)}|^2$ on $N$ can be either monotonic or oscillatory depending on the parameters of the fiber, the grating, and the light field. The reason for the oscillatory behavior is that the parameter $\theta_{\sigma}$ is, in general, a complex number, and therefore the hyperbolic functions $\sinh(N\theta_{\sigma})$ and $\cosh(N\theta_{\sigma})$ in expressions (\ref{f25}) may become oscillatory functions. In the case of Fig. \ref{fig2}(a), 
$|R_y^{(N)}|^2$ (dashed line) dominates $|R_x^{(N)}|^2$ (solid line). In the case of Fig. \ref{fig2}(c), $|R_x^{(N)}|^2$ (solid line) dominates $|R_y^{(N)}|^2$ (dashed line) when $N$ is large enough. In the case of Fig. \ref{fig2}(b), both $|R_x^{(N)}|^2$ (solid line) and $|R_y^{(N)}|^2$ (dashed line) achieve similar significant values when $N$ is around 200.

%%%%%%%%%%%%%%%%%%%%%%% Figure 3

%%%%%%%%%%%%%%%%%%%%%%% Figure 4

%%%%%%%%%%%%%%%%%%%%%%% Figure 5

%%%%%%%%%%%%%%%%%%%%%%% Figure 6

Figures \ref{fig3}--\ref{fig6} show that the reflectivity $|R_{\sigma}^{(N)}|^2$ of the grating oscillates with increasing $\lambda_G$, $d$, $h$, or $\lambda$. We observe that $|R_{\sigma}^{(N)}|^2$ can achieve significant values when the parameters $\lambda_G$, $d$, $h$, or $\lambda$ are appropriate. Figures \ref{fig3}--\ref{fig6} show that the region where $|R_x^{(N)}|^2$ achieves significant values is different from the region where $|R_y^{(N)}|^2$ does so. 
The magnitude and width of the maximum of $|R_y^{(N)}|^2$ are, in general, larger than those of the maximum of $|R_x^{(N)}|^2$. Figures \ref{fig3}(c), \ref{fig4}(c), \ref{fig5}(c), and \ref{fig6}(c) show that, for $N=240$, both $|R_x^{(N)}|^2$ and $|R_y^{(N)}|^2$ can achieve similar significant values when $\lambda_G=364.5$ nm, $d=100$ nm, $h=150$ nm, and $\lambda=852$ nm. 

We recognize that the length $h$ and the depth $d$ of the holes used in the above numerical results 
are substantial fractions of the fiber radius. For such a grating, the losses to radiation modes are not small. Therefore, our analytical and numerical results are not quantitatively valid. However, they can still be used to qualitatively understand the underlying physics and to explain many features of the nanofiber cavities produced by the focused-ion-beam milling method \cite{nanogrooves}.

\section{Nanofiber cavity}
\label{sec:cavity}

\begin{figure}[tbh]
\begin{center}
 \includegraphics{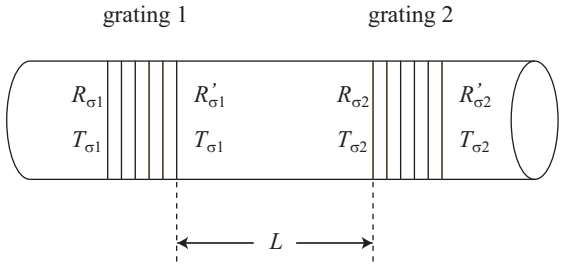}
 \end{center}
\caption{Cavity formed by a nanofiber with two Bragg gratings.
}
\label{fig7}
\end{figure}

We consider a cavity formed by a nanofiber with two Bragg gratings (see Fig. \ref{fig7}). We label the gratings by the index $j$, where $j=1,2$. Each of the gratings can be either a single grating or a set of multiple gratings. In the case of a multiple grating, the reflection coefficients for the left and right sides of the grating may differ from each other. Therefore, to be general, we denote the reflection coefficient for a light field with a principal polarization $\sigma=x,y$, incident onto 
the left or right side of grating $j=1,2$ by the notation $R_{\sigma j}$ or $R'_{\sigma j}$, respectively.
The transmission coefficient for grating $j$ is $T_{\sigma j}$ and does not depend on the incidence side of the grating. The distance between the two gratings of the cavity is denoted by $L$. When we use ray optics and combine multiply reflected beams, we can easily calculate the reflection and transmission coefficients of the set of the two gratings, that is, for the whole nanofiber cavity.

%%%%%%%%%%%%%%%%%%%%%%% Figure 7

The reflection coefficients $R_{\sigma}$ and $R'_{\sigma}$ for the left and right sides, respectively, of the set of the two gratings are found to be
\begin{eqnarray}
R_{\sigma}&=&R_{\sigma 1}+\frac{T_{\sigma 1}^2R_{\sigma 2}e^{2i\beta L}}{1-R'_{\sigma 1}R_{\sigma 2}e^{2i\beta L}},\nonumber\\
R'_{\sigma}&=&R'_{\sigma 2}+\frac{T_{\sigma 2}^2R'_{\sigma 1}e^{2i\beta L}}{1-R'_{\sigma 1}R_{\sigma 2}e^{2i\beta L}}.
\label{f27}
\end{eqnarray}
Meanwhile, the transmission coefficient $T_{\sigma}$ for the set of the two gratings is found to be
\begin{equation}
T_{\sigma}=\frac{T_{\sigma 1}T_{\sigma 2}e^{i\beta L}}{1-R'_{\sigma 1}R_{\sigma 2}e^{2i\beta L}}.
\label{f28}
\end{equation}
Hence, the transmission of the cavity formed by the two gratings is
\begin{equation}
|T_{\sigma}|^2=\frac{|T_{\sigma 1}|^2|T_{\sigma 2}|^2}{(1-|R_{\sigma 1}||R_{\sigma 2}|)^2+4|R_{\sigma 1}||R_{\sigma 2}|\sin^2\Theta_{\sigma}},
\label{f29}
\end{equation}
where $\Theta_{\sigma}=(\phi'_{\sigma 1}+\phi_{\sigma 2})/2+\beta L$, with $\phi'_{\sigma 1}$ and $\phi_{\sigma 2}$ being the phases of the complex reflection coefficients $R'_{\sigma 1}$ and $R_{\sigma 2}$, respectively. 

We consider the case where the separation between the two gratings of the cavity is much larger than the light wavelength, that is, $L\gg\lambda$. We assume that the frequency $\omega$ of the light field is in a small interval around a central frequency $\omega_0$. In this case, we can use the expansion $\beta(\omega)=\beta_0+\delta/v_g$, where $\beta_0=\beta(\omega_0)$ is the central value of the fiber propagation constant, $\delta=\omega-\omega_0$ is the detuning of the field, and $v_g=(d\beta/d\omega)^{-1}|_{\omega=\omega_0}$ is the group velocity of the guided light field. Then, we have 
\begin{equation} 
\Theta_{\sigma}=(\phi'_{\sigma 1}+\phi_{\sigma 2})/2+\beta_0L+L\delta/v_g.
\label{f29a}
\end{equation} 
When the separation $L$ between the two gratings is sufficiently large as compared to their thicknesses, the moduli and phases of the complex reflection and transmission coefficients of the gratings vary very little with the frequency in the range of a few times of $\pi v_g/L$ around $\omega_0$. Then, we find from Eqs. (\ref{f29}) and (\ref{f29a}) that the separation between the resonances is $\Delta\omega_{\mathrm{res}}=\pi v_g/L$ and the width of the resonances is $\delta\omega_{\mathrm{res}}=(1-|R_{\sigma 1}||R_{\sigma 2}|)v_g/(\sqrt{|R_{\sigma 1}||R_{\sigma 2}|}L)$. Hence, the finesse of the cavity is estimated to be $F_{\sigma}\equiv\Delta\omega_{\mathrm{res}}/\delta\omega_{\mathrm{res}}=\pi\sqrt{|R_{\sigma 1}||R_{\sigma 2}|}/(1-|R_{\sigma 1}||R_{\sigma 2}|)$. It is clear that the cavity transmission $|T_{\sigma}|^2$ and the cavity finesse $F_{\sigma}$ depend on the principal polarization $\sigma=x,y$ of the field. 

%%%%%%%%%%%%%%%%%%%%%%% Figure 8
\begin{figure}[tbh]
\begin{center}
 \includegraphics{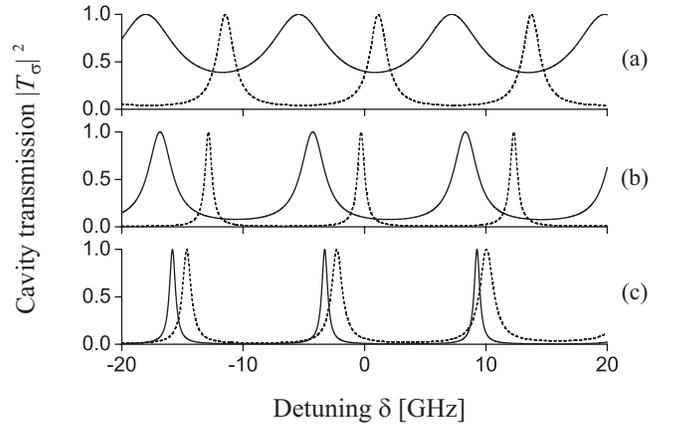}
 \end{center}
\caption{Cavity transmission $|T_{\sigma}|^2$ as a function of the detuning $\delta$. The solid and dashed curves are for the principal $x$ and $y$ polarizations, respectively. 
The cavity length is $L=7.6$ mm. The two FBG mirrors are identical. 
The number of steps in each mirror is $N=60$ (a), 120 (b), and 240 (c). The grating period is $\lambda_G$ =364.5 nm. Other parameters are as in Fig. \ref{fig2}.
}
\label{fig8}
\end{figure}

We plot in Fig. \ref{fig8} the cavity transmission $|T_{\sigma}|^2$ as a function of the detuning $\delta$. Comparison between the solid curves (for the $x$ polarization) and the dashed curves (for the $y$ polarization) show that the positions and widths of the resonances are different for the different principal $x$ and $y$ polarizations of the light field.

%%%%%%%%%%%%%%%%%%%%%%% Figure 9
\begin{figure}[tbh]
\begin{center}
 \includegraphics{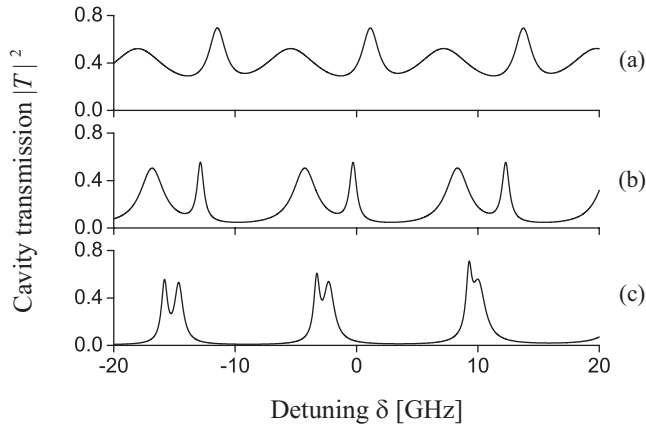}
 \end{center}
\caption{Tuning profile of the cavity transmission $|T|^2=|T_x|^2\cos^2\varphi+|T_y|^2\sin^2\varphi$ for $\varphi=\pi/4$. 
Other parameters are as for Fig. \ref{fig8}.
}
\label{fig9}
\end{figure}

We emphasize that Eqs. (\ref{f27})--(\ref{f29}) are valid only when the polarization of the light field is linear and is oriented along the principal symmetry axis $x$ or $y$. When the polarization of the light field is linear and is oriented at an angle $\varphi$ in the $xy$ plane, the input field can be presented in the form $\mathbf{E}_{\mathrm{in}}=A(\boldsymbol{\mathcal{E}}_x\cos\varphi+\boldsymbol{\mathcal{E}}_y\sin\varphi)$, where $\boldsymbol{\mathcal{E}}_x=(\boldsymbol{\mathcal{E}}_{1}+\boldsymbol{\mathcal{E}}_{-1})/\sqrt2$ and $\boldsymbol{\mathcal{E}}_y=(\boldsymbol{\mathcal{E}}_{1}-\boldsymbol{\mathcal{E}}_{-1})/i\sqrt2$ are the profile functions of the guided modes with the principal $x$ and $y$ polarizations, respectively, and $A$ characterizes the field strength. Then, the 
reflected and transmitted fields are found to be 
\begin{eqnarray}
\mathbf{E}_{\mathrm{rf}}&=&A(\boldsymbol{\mathcal{E}}_xR_x\cos\varphi+\boldsymbol{\mathcal{E}}_yR_y\sin\varphi),
\nonumber\\
\mathbf{E}_{\mathrm{tr}}&=&A(\boldsymbol{\mathcal{E}}_xT_x\cos\varphi+\boldsymbol{\mathcal{E}}_yT_y\sin\varphi).
\label{f30}
\end{eqnarray}
Equations (\ref{f30}) show that, when $R_x\not=R_y$ and $T_x\not=T_y$, the directions of the polarization vectors of the input, reflected, and transmitted fields are, in general, different from each other. Consequently, the polarization vector of the field may be rotated after being reflected from or transmitted through the cavity. Such a rotation of the polarization vector of the field is a result of the breaking of the cylindrical symmetry of the fiber. The rotation vanishes when
$\varphi=0$ or $\pi/2$, that is, when the polarization vector of the light field is oriented along the principal axis $x$ or $y$.

Due to the mode orthogonality, the power of the field is the sum of the powers of the individual $x$- and $y$-polarized modes. Consequently, the reflectivity and transmission of the field in terms of the power are given by
\begin{eqnarray}
|R|^2&=&|R_x|^2\cos^2\varphi+|R_y|^2\sin^2\varphi,
\nonumber\\
|T|^2&=&|T_x|^2\cos^2\varphi+|T_y|^2\sin^2\varphi.
\label{f31}
\end{eqnarray} 
In general, the tuning profile of the cavity transmission $|T|^2$ contains the features
of both the profiles of $|T_x|^2$ and $|T_y|^2$. We note that the discussions around Eqs. (\ref{f30}) and (\ref{f31}) are valid not only for the case of a cavity but also for the case of a single grating.

We plot in Fig. \ref{fig9} the tuning profile of the cavity transmission $|T|^2=|T_x|^2\cos^2\varphi+|T_y|^2\sin^2\varphi$ for the angle $\varphi=\pi/4$ of the polarization orientation of the light field. 
All other parameters are the same as those for Fig. \ref{fig8}. 
The figure shows clearly that the profile of $|T|^2$ has double-peak structure. In such  structure, one peak corresponds to the resonance in the profile of $|T_x|^2$ and the other peak corresponds to the resonance in the profile of $|T_y|^2$. Comparison between parts (a), (b), and (c) of the figure shows that 
the tuning profile of the transmission of the cavity depends on the number of holes. 

%%%%%%%%%%%%%%%%%%%%%%% Figure 10
\begin{figure}[tbh]
\begin{center}
 \includegraphics{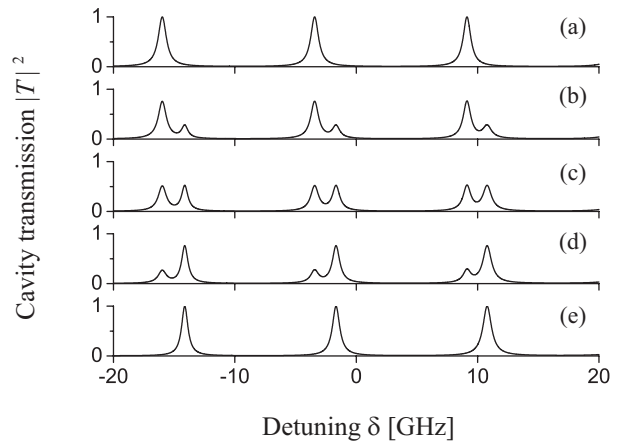}
 \end{center}
\caption{Tuning profile of the cavity transmission $|T|^2=|T_x|^2\cos^2\varphi+|T_y|^2\sin^2\varphi$ for different values $\varphi=0$ (a), $\pi/6$ (b), $\pi/4$ (c), $\pi/3$ (d), and $\pi/2$ (e) of the orientation angle of the polarization vector of the light field. The number of steps is $N=210$. Other parameters are as for Fig. \ref{fig8}.
}
\label{fig10}
\end{figure}

We plot in Fig. \ref{fig10} the tuning profile of the cavity transmission $|T|^2=|T_x|^2\cos^2\varphi+|T_y|^2\sin^2\varphi$ for different values of the orientation angle $\varphi$ of the polarization vector of the light field in the $xy$ plane. It is clear from the figure that the tuning profile of $|T|^2$ substantially depends on the angle $\varphi$. The profile has a single peak when $\varphi=0$ (a) or $\pi/2$ (e), that is, when the polarization vector of the light field is oriented along the principal axis $x$ or $y$, respectively. In the cases of parts (b), (c), and (d), where $\varphi=\pi/6$, $\pi/4$, and $\pi/3$, respectively, the profile has double-peak structure. It is interesting to note that the two peaks have almost the same width. The reason is that, for the chosen parameters, the different principal $x$ and $y$ polarizations of the light field correspond to almost the same reflectivity of the FBG mirrors and, consequently, to almost the same finesse of the cavity.

%%%%%%%%%%%%%%%%%%%%%%% Figure 11
\begin{figure}[tbh]
\begin{center}
 \includegraphics{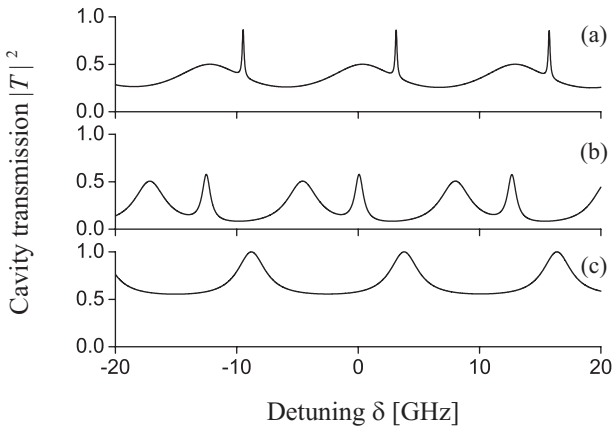}
 \end{center}
\caption{Tuning profile of the cavity transmission $|T|^2=|T_x|^2\cos^2\varphi+|T_y|^2\sin^2\varphi$ for the grating period $\lambda_G=363$ nm (a), 364.4 nm (b), and 366 nm (c).
The orientation angle of the field polarization vector is $\varphi=\pi/4$. The number of steps is $N=100$. Other parameters are as for Fig. \ref{fig8}.
}
\label{fig11}
\end{figure}

We plot in Fig. \ref{fig11} the tuning profile of the cavity transmission $|T|^2$ for different values of the grating period $\lambda_G$. It is clear from the figure that the tuning profile of $|T|^2$ is very sensitive to the magnitude of the grating period $\lambda_G$. In the case of part (a), where $\lambda_G=363$ nm, the mirror reflectivity $|R_y|^2\cong0.96$ and cavity finesse $F_y\cong70$ for the $y$ polarization are large as compared to the corresponding values $|R_x|^2\cong0.16$ and $F_x\cong1.5$ for the $x$ polarization. This is the reason why a narrow peak appears on top of a broad peak. In the case of part (c), where $\lambda_G=366$ nm, the mirror reflectivity and cavity finesse for the $y$ polarization are negligible. Therefore, we observe in this case a significant background in the tuning profile of the cavity transmission $|T|^2$.

\section{Summary}
\label{sec:summary} 

In conclusion, we have studied nanofibers with Bragg gratings from equidistant holes. We have calculated analytically and numerically the reflection and transmission coefficients for a single grating and also for a cavity formed by two gratings. We have shown that the reflection and transmission coefficients of the gratings substantially depend on the number of holes, the hole length, the hole depth, the grating period, and the light wavelength. We have found that the reflection and transmission coefficients of the gratings depend on the orientation of the polarization vector of light with respect to the holes. Such a dependence is a result of the fact that the cross section of the gratings is not cylindrically symmetric. Due to this property, the polarization vector of the field may be rotated after being reflected from or transmitted through the gratings. In addition, the tuning profile of the cavity transmission may have double-peak structure, whose components result from the resonances of the field with the different principal $x$ and $y$ polarizations. We emphasize that the only approximation used in our treatment is the omission of the radiation modes from the coupled-mode equations. This approximation is valid when the losses to the radiation modes are not serious, that is, when the characteristic size of the holes is small as compared to the fiber radius. It is interesting to note that, even when the length and depth of the holes are substantial fractions of the fiber radius, our analytical and numerical results are in good qualitative agreement with the experimental results \cite{nanogrooves}.

\begin{acknowledgments}
We acknowledge useful discussions with M. Morinaga.
\end{acknowledgments}

\appendix

\section{Mode functions of the fundamental guided modes of a nanofiber}
\label{sec:guided}

For the fundamental guided modes, the propagation constant $\beta$ is determined by the
fiber eigenvalue equation \cite{fiber books}
\begin{eqnarray}
\frac{J_0(h a)}{h a J_1(h a)}&=&
-\frac{n_1^2+n_2^2}{2n_1^2}\frac{K_1'(q a)}{q a K_1(q a)}+ \frac{1}{h^2 a^2}
\nonumber\\&&\mbox{}
-\Bigg[\left(\frac{n_1^2-n_2^2}{2n_1^2}\frac{K_1'(q a)}{q a K_1(q a)}\right)^2
\nonumber\\&&\mbox{}
+\frac{\beta^2}{n_1^2 k^2}\left(\frac{1}{q^2a^2}+\frac{1}{h^2a^2}\right)^2 \Bigg]^{1/2}.
\label{G1}
\end{eqnarray}
Here the parameters $h=(n_1^2k^2-\beta^2)^{1/2}$ and $q=(\beta^2-n_2^2k^2)^{1/2}$ characterize the fields inside and outside the fiber, respectively. The notations $J_n$ and $K_n$ stand for the Bessel functions of the first kind and the modified Bessel functions of the second kind, respectively. 

We present the mode functions of the electric and magnetic parts of the fundamental guided mode $\mu=(\omega, f, l)$ in the forms $\boldsymbol{\mathcal{E}}_{\mu}=\mathbf{e}^{(\mu)}e^{il\varphi}$ and $\boldsymbol{\mathcal{H}}_{\mu}=\mathbf{h}^{(\mu)}e^{il\varphi}$, respectively \cite{fiber books}. The
cylindrical components of the phase-shifted electric-part guided-mode function $\mathbf{e}^{(\mu)}$ are given, for $r<a$, by
\begin{eqnarray}
e_{r}^{(\mu)}&=&iC\frac{q}{h}\frac{K_1(qa)}{J_1(ha)}[(1-s)J_0(hr)-(1+s)J_2(hr) ],
\nonumber\\
e_{\varphi}^{(\mu)}&=&-lC\frac{q}{h}\frac{K_1(qa)}{J_1(ha)}[(1-s)J_0(hr)+(1+s)J_2(hr) ],
\nonumber\\
e_{z}^{(\mu)}&=& fC\frac{2q}{\beta}\frac{K_1(qa)}{J_1(ha)}J_1(hr),
\label{G2}
\end{eqnarray}
and, for $r>a$, by
\begin{eqnarray}
e_{r}^{(\mu)}&=&iC[(1-s)K_0(qr)+(1+s)K_2(qr) ],
\nonumber\\
e_{\varphi}^{(\mu)}&=&-lC[(1-s)K_0(qr)-(1+s)K_2(qr) ],
\nonumber\\
e_{z}^{(\mu)}&=& fC\frac{2q}{\beta}K_1(qr).
\label{G3}
\end{eqnarray} 
Here the parameter $s$ is defined as 
$s=({1}/{q^2a^2}+{1}/{h^2a^2})/[{J_1^\prime (ha)}/{haJ_1(ha)}+{K_1^\prime (qa)}/{qaK_1(qa)}]$.
Similarly, the cylindrical components of the phase-shifted magnetic-part guided-mode function $\mathbf{h}^{(\mu)}$ are given, for $r<a$, by
\begin{eqnarray}
h_r^{(\mu)}&=&flC\frac{\omega\epsilon_0 n_1^2q}{\beta h}\frac{K_1(qa)}{J_1(ha)}
[(1-s_1)J_0(hr)
\nonumber\\&&\mbox{}
+(1+s_1)J_2(hr)],
\nonumber\\
h_{\varphi}^{(\mu)}&=&ifC\frac{\omega\epsilon_0 n_1^2q}{\beta h}\frac{K_1(qa)}{J_1(ha)}
[(1-s_1)J_0(hr)
\nonumber\\&&\mbox{}
-(1+s_1)J_2(hr)],
\nonumber\\
h_z^{(\mu)}&=& ilC\frac{2q}{\omega\mu_0} s \frac{K_1(qa)}{J_1(ha)} J_1(hr),
\label{G4}
\end{eqnarray} 
and, for $r>a$, by 
\begin{eqnarray}
h_r^{(\mu)}&=&flC\frac{\omega\epsilon_0 n_2^2}{\beta}
[(1-s_2)K_0(qr)-(1+s_2)K_2(qr)],
\nonumber\\
h_{\varphi}^{(\mu)}&=&ifC\frac{\omega\epsilon_0 n_2^2}{\beta}
[(1-s_2)K_0(qr)+(1+s_2)K_2(qr)],
\nonumber\\
h_z^{(\mu)}&=&ilC\frac{2q}{\omega\mu_0}s K_1(qr).
\label{G5}
\end{eqnarray}

The power of the field in the fundamental guided mode $\mu$ is given by
\begin{equation}
P_{\mu}=\frac{1}{2} \int \mathrm{Re}[\mathbf{u}_z\cdot (\mathbf{e}^{(\mu)}\times \mathbf{h}^{(\mu)*})] \,d\mathbf{r}.
\label{G7}
\end{equation}
It can be shown that
\begin{eqnarray}
P_{\mu}&=&|C|^2\frac{\pi a^2\omega\epsilon_0 n_1^2}{\beta}\frac{q^2K_1^2(qa)}{h^2J_1^2(ha)}
\{(1-s)(1-s_1)
\nonumber\\&&\mbox{}\times
[J_0^2(ha)+J_1^2(ha)]+(1+s)(1+s_1)
\nonumber\\&&\mbox{}\times
[J_2^2(ha)-J_1(ha)J_3(ha)]\}
\nonumber\\&&\mbox{}
+|C|^2\frac{\pi a^2\omega\epsilon_0 n_2^2}{\beta}
\{(1-s)(1-s_2)
\nonumber\\&&\mbox{}\times
[K_1^2(qa)-K_0^2(qa)]+(1+s)(1+s_2)
\nonumber\\&&\mbox{}\times
[K_1(qa)K_3(qa)-K_2^2(qa)]\}.
\label{G8}
\end{eqnarray}

In the case where the coefficient $C$ is real, we have the following symmetry relations: 
\begin{eqnarray}\label{G9}
e_r^{(\omega,f,l)}&=&e_r^{(\omega,-f,l)}=e_r^{(\omega,f,-l)},\nonumber\\
e_{\varphi}^{(\omega,f,l)}&=&e_{\varphi}^{(\omega,-f,l)}=-e_{\varphi}^{(\omega,f,-l)},\nonumber\\
e_z^{(\omega,f,l)}&=&-e_z^{(\omega,-f,l)}=e_z^{(\omega,f,-l)},
\end{eqnarray}
and
\begin{equation}\label{G10}
e_r^{(\mu)*}=-e_r^{(\mu)},\quad
e_\varphi^{(\mu)*}=e_\varphi^{(\mu)},\quad
e_z^{(\mu)*}=e_z^{(\mu)}.
\end{equation}


\begin{thebibliography}{99}

\bibitem{Hill} K. O. Hill, Y. Fujii, D. C. Johnson, and B. S. Kawasaki, ``Photosensitivity in optical fiber waveguides: Application to reflection filter fabrication,''
Appl. Phys. Lett. \textbf{32}, 647--649 (1978).

\bibitem{Meltz} G. Meltz, W. W. Morey, and W. H. Glenn, ``Formation of Bragg gratings in optical fibers by a transverse holographic method,'' Opt. Lett. \textbf{14}, 823--825 (1989).

\bibitem{Kashyap} R. Kashyap, \textit{Fiber Bragg Gratings} (Academic, New York, 1999).

\bibitem{Canning} For a recent review see, for example, J. Canning, ``Fiber gratings and devices for sensors and lasers,'' Lasers and Photonic Rev. \textbf{2}, 275--289 (2008).

\bibitem{Wan} X. Wan and H. F. Taylor, ``Intrinsic fiber Fabry-Perot temperature sensor with fiber Bragg grating mirrors,'' Opt. Lett. \textbf{27}, 1388--1390 (2002).

\bibitem{Chow} J. H. Chow, I. C. M. Littler, G. de Vine, D. E. McClelland, and M. B. Gray, ``Phase-sensitive interrogation of fiber Bragg grating resonators for sensing applications,'' 
 J. Lightwave Technol. \textbf{23}, 1881--1889 (2005). 

\bibitem{Gupta} M. Gupta, H. Jiao, and A. O'Keefe, 
``Cavity-enhanced spectroscopy in optical fibers,'' Opt. Lett. \textbf{27}, 1878--1880 (2002).

\bibitem{Mazur's Nature} L. Tong, R. R. Gattass, J. B. Ashcom, S. He, J. Lou, M. Shen, I. Maxwell, and E. Mazur, ``Subwavelength-diameter silica wires for low-loss optical wave guiding,'' 
 Nature (London) \textbf{426}, 816--819 (2003).

\bibitem{Birks} T. A. Birks, W. J. Wadsworth, and P. St. J. Russell, ``Supercontinuum generation in tapered fibers,'' Opt. Lett. \textbf{25}, 1415--1417 (2000). 

\bibitem{taper} J. C. Knight, G. Cheung, F. Jacques, and T. A. Birks, ``Phase-matched excitation of whispering-gallery-mode resonances by a fiber taper,'' Opt. Lett. \textbf{22}, 1129--1131 (1997).

\bibitem{Dowling} J. P. Dowling and J. Gea-Banacloche, ``Evanescent light-wave atom mirrors, resonators, waveguides, and traps,'' Adv. At., Mol., Opt. Phys. \textbf{37}, 1--94 (1996).

\bibitem{onecolor} V. I. Balykin, K. Hakuta, Fam Le Kien, J. Q. Liang, and M. Morinaga, ``Atom trapping and guiding with a subwavelength-diameter optical fiber,'' Phys. Rev. A \textbf{70}, 011401(R) (2004). 

\bibitem{twocolors} Fam Le Kien, V. I. Balykin, and K. Hakuta, ``Atom trap and waveguide using a two-color evanescent light field around a subwavelength-diameter optical fiber,'' 
 Phys. Rev. A \textbf{70}, 063403 (2004).

\bibitem{Rauschenbeutel} G. Sagu\'{e}, A. Baade, and A. Rauschenbeutel, ``Blue-detuned evanescent field surface traps for neutral atoms based on mode interference in ultrathin optical fibres,'' New J. Phys. \textbf{10}, 113008 (2008).

\bibitem{twocolor experiment} E. Vetsch, D. Reitz, G. Sagu\'{e}, R. Schmidt, S. T. Dawkins, and A. Rauschenbeutel, ``Optical interface created by laser-cooled atoms trapped in the evanescent field surrounding an optical nanofiber,'' Phys. Rev. Lett. \textbf{104}, 203603 (2010).

\bibitem{Bures and Ghosh} J. Bures and R. Ghosh, ``Power density of the evanescent field in the vicinity of a tapered fiber,'' J. Opt. Soc. Am. A \textbf{16}, 1992--1996 (1999).

\bibitem{zero mode} M.J. Levene, J. Korlach, S.W. Turner, M. Foquet, H.G. Craighead, and W.W. Webb, ``Zero-mode waveguides for single-molecule analysis at high concentrations,'' Science \textbf{299}, 682--686 (2003).

\bibitem{Kali} K. P. Nayak, P. N. Melentiev, M. Morinaga, Fam Le Kien, V. I. Balykin, and K. Hakuta, ``Optical nanofiber as an efficient tool for manipulating and probing atomic fluorescence,'' Opt. Express \textbf{15}, 5431--5438 (2007). 

\bibitem{Kali antibunching} K. P. Nayak, Fam Le Kien, M. Morinaga, and K. Hakuta, ``Antibunching and bunching of photons in resonance fluorescence from a few atoms into guided modes of an optical nanofiber,'' Phys. Rev. A \textbf{79}, 021801(R) (2009).

\bibitem{fibercavity} Fam Le Kien and K. Hakuta, ``Intracavity electromagnetically induced transparency in atoms around a nanofiber with a pair of Bragg grating mirrors,'' Phys. Rev. A \textbf{79}, 043813 (2009).

\bibitem{Mabuchi} H. Mabuchi, Q. A. Turchette, M. S. Chapman, and H. J. Kimble, ``Real-time detection of individual atoms falling through a high-finesse optical cavity,'' Opt. Lett. \textbf{21}, 1393--1395 (1996).

\bibitem{Hood Science} C. J. Hood, T. W. Lynn, A. C. Doherty, A. S. Parkins, and H. J. Kimble, ``The atom-cavity microscope: Single atoms bound in orbit by single photons,'' Science \textbf{287}, 1447--1453 (2000).

\bibitem{Aoki} T. Aoki, B. Dayan, E. Wilcut, W. P. Bowen, A. S. Parkins, T. J. Kippenberg, K. J. Vahala, and H. J. Kimble, ``Observation of strong coupling between one atom and a monolithic microresonator,'' Nature (London) \textbf{443}, 671--674 (2006).

\bibitem{Lukin1998} M. D. Lukin, M. Fleischhauer, M. O. Scully, and V. L. Velichansky, ``Intracavity electromagnetically induced transparency,'' Opt. Lett. \textbf{23}, 295--297 (1998).

\bibitem{Xiao} H. Wang, D. J. Goorskey, W. H. Burkett, and M. Xiao, ``Cavity-linewidth narrowing by means of electromagnetically induced transparency,'' Opt. Lett. \textbf{25}, 1732--1734 (2000).

\bibitem{Zhu} J. Zhang, G. Hernandez, and Y. Zhu, ``Slow light with cavity electromagnetically induced transparency,'' Opt. Lett. \textbf{33}, 46--48 (2008).

\bibitem{cavityspon} Fam Le Kien and K. Hakuta, ``Cavity-enhanced channeling of emission from an atom into a nanofiber,'' Phys. Rev. A \textbf{80}, 053826 (2009).

\bibitem{cavitytrap} Fam Le Kien and K. Hakuta, ``Effect of an atom on a quantum guided field in a weakly driven fiber-Bragg-grating cavity,'' Phys. Rev. A \textbf{81}, 023812 (2010).

\bibitem{nanogrooves} K. P. Nayak, K. Nakajima, Fam Le Kien, H. T. Miyazaki, Y. Sugimoto, and K. Hakuta, ``Optical nanofiber cavity: A novel workbench for cavity QED,'' in
\textit{Frontiers in Optics 2010/Laser Science XXVI Conference}, Technical Digest (Optical Society of America, 2010), paper FTuT5.

\bibitem{FIB} V. Hodzic, J. Orloff, and C.C. Davis, ``Periodic structures on biconically tapered optical fibers using ion beam milling and boron implantation,'' J. Lightwave Technol. \textbf{22}, 1610--1614 (2004).

\bibitem{fiber books} See, for example, 
D. Marcuse, \textit{Light Transmission Optics} 
(Krieger, Malabar, FL, 1989);
A. W. Snyder and J. D. Love, \textit{Optical Waveguide Theory} (Chapman and Hall, New York, 1983).

\end{thebibliography}
\end{document}